\documentclass[11pt]{elsart}
\usepackage{amssymb}
\usepackage[reqno]{amsmath}

\usepackage{graphicx}
\usepackage{subfigure}

\numberwithin{equation}{section}

\begin{document}

\input epsf

\begin{flushright}
\baselineskip=12pt
CERN--TH/98--93,\quad UTPT--98--04
\end{flushright}

\begin{frontmatter}

\title{The dynamical stability of the static real scalar field solutions to the {E}instein-{K}lein-{G}ordon equations revisited}
\author[CERN]{M.~A. Clayton\thanksref{mike}},
\author[Toronto]{L. Demopoulos\thanksref{terry}} and
\author[Toronto]{J. L\'{e}gar\'{e}\thanksref{jacques}}

\address[CERN]{CERN--Th, CH--1211 Geneva 23, Switzerland}
\address[Toronto]{Department of physics, University of Toronto, Toronto, \textsc{on}, Canada M5S~1A7}

\thanks[mike]{E-mail: \texttt{Michael.A.Clayton@cern.ch}}
\thanks[terry]{E-mail: \texttt{terry@medb.physics.utoronto.ca}}
\thanks[jacques]{E-mail: \texttt{jacques@medb.physics.utoronto.ca}}

\begin{abstract}
We re-examine the dynamical stability of the nakedly singular, static, spherically symmetric solutions of the Einstein-Klein Gordon system.
We correct an earlier proof of the instability of these solutions, and demonstrate that there are solutions to the massive Klein-Gordon system that are perturbatively stable.
\end{abstract}

\end{frontmatter}

\section{Introduction}
\label{sect:Intro}

We are re-considering the dynamical stability of the non-trivial static, spherically symmetric solutions of a real, self-coupled, self-gravitating scalar field.
The case where the scalar field is massless and the potential vanishes has analytically known static solutions~\cite{Wyman:1981} which, as claimed by Jetzer and Scialom~\cite{Jetzer+Scialom:1992}, are unstable against linear perturbations of the system.
This is in accordance with the cosmic censorship hypothesis~\cite{Wald:1997} and the results of Christodoulou's study of this system (beginning with~\cite{Christodoulou:1986}) which led to the conclusion that the set of initial data that collapse to these nakedly singular solutions is of measure zero~\cite{Christodoulou:1994}.

That the massive scalar field has quite a different flavour is indicated by the presence of periodic in time, soliton solutions~\cite{Seidel+Suen:1991} and the richer structure recently discovered in collapsing solutions that are near the threshold of black hole formation~\cite{Brady+Chambers+Goncalves:1997}.
As one further piece of the puzzle, we show here that there are nakedly singular solutions to the massive scalar field equations which are stable against spherically symmetric perturbations.
This is in contradiction to the results reported by Jetzer and Scialom~\cite{Jetzer+Scialom:1992}; however, their calculations contain a flaw that leads to the erroneous conclusion that all of these nakedly singular solutions are perturbatively unstable.
The results from the corrected version of their variational bound are consistent with the explicit determination of the perturbative mode.

The system of equations that we are considering is straightforward to derive.
We will state here the basic equations and conventions we use.
We consider spherically symmetric spacetimes with the metric in the diagonal form ($\nu=\nu(t,r)$ and $\lambda=\lambda(t,r)$)
\begin{equation}
\mathrm{g}=\mathrm{diag}
(\mathrm{e}^\nu,-\mathrm{e}^\lambda,-r^2,-r^2\sin^2(\theta)),
\end{equation}
dynamically coupled to a scalar field with Lagrangian
\begin{subequations}
\begin{equation}
\mathcal{L}=\tfrac{1}{2}\mathrm{g}^{\mu\nu}
\partial_\mu[\phi]\partial_\nu[\phi]
-V[\phi],
\end{equation}
where the potential and its variation are given by
\begin{equation}
V[\phi]=\tfrac{1}{2}m^2\phi^2+\tfrac{1}{4}\kappa\phi^4,\quad
\delta_\phi V[\phi]=m^2\phi+\kappa\phi^3.
\end{equation}
\end{subequations}
Note that we have chosen $16\pi\mathrm{G}=\mathrm{c}=1$ and $\phi$ is taken to be dimensionless;  this implies that $m$ has dimension of an an inverse length and $\kappa$ an inverse length squared.

Einstein's equations may be written $R_{\mu\nu}=\tfrac{1}{2}(\partial_\mu[\phi]\partial_\nu[\phi]-\mathrm{g}_{\mu\nu}V[\phi])$, from which we find
\begin{subequations}\label{eq:feq used}
\begin{align}
R_{01}&=\tfrac{1}{r}\partial_t[\lambda]
=\tfrac{1}{2}\partial_t[\phi]\partial_r[\phi],\\
\mathrm{e}^{-\nu}R_{00}+\mathrm{e}^{-\lambda}R_{11}&=
\tfrac{1}{r}\mathrm{e}^{-\lambda}\partial_r[\nu+\lambda]
=\tfrac{1}{2}\mathrm{e}^{-\nu}(\partial_t[\phi])^2
+\tfrac{1}{2}\mathrm{e}^{-\lambda}(\partial_r[\phi])^2,\\
R_{22}&=-\tfrac{r}{2}\mathrm{e}^{-\lambda}\partial_r[\nu-\lambda]
+1-\mathrm{e}^{-\lambda}
=\tfrac{1}{2}r^2V[\phi].
\end{align}
The variation of the matter action with respect to $\phi$ results in the wave equation 
\begin{multline}
\mathrm{e}^{-\nu}\partial_t^2[\phi]
-\tfrac{1}{2}\mathrm{e}^{-\nu}\partial_t[\nu-\lambda]\partial_t[\phi]\\
-\mathrm{e}^{-\lambda}\partial_r^2[\phi]
-\mathrm{e}^{-\lambda}
(\tfrac{2}{r}+\tfrac{1}{2}\partial_r[\nu-\lambda])\partial_r[\phi]
+\delta_\phi V[\phi]=0.
\end{multline}
\end{subequations}

In Section~\ref{sect:static}, we present the numerical solutions of the static limit of~\eqref{eq:feq used}, with the Appendix devoted to the analytic solution of the static equations with vanishing potential.
In Section~\ref{sect:instability} we set up the perturbative stability problem and present the numerically generated ground state energies of the perturbative mode, indicating that the spacetimes with vanishing potential are unstable but that there is a branch of the spacetimes with non-vanishing scalar field mass which is perturbatively stable.
In Section~\ref{sect:var}, we examine the variational method introduced by Jetzer and Scialom to put an upper bound on the ground state energy, showing that the corrected method is consistent with the results of Section~\ref{sect:instability}.

\section{The Static Background Spacetimes}
\label{sect:static}

Throughout, we will make use of the dimensionless radial coordinate scaled by the asymptotic Schwarzschild mass parameter $M_{\textsc{s}}$ of the system: $x:=r/(2M_{\textsc{s}})$.
In the case of a non-zero inverse length scale $m$ or $\sqrt{\kappa}$ we will define the dimensionless quantities $\tilde{m}:=2M_{\textsc{s}}m$ and $\tilde{\kappa}:=4M^2_{\textsc{s}}\kappa$.
We then re-write the static limit of~\eqref{eq:feq used} in a form appropriate for the numerical integration implemented below:
\begin{subequations}\label{eq:static eqns}
\begin{gather}
\label{eq:st lambda}
\partial_x[\mathrm{e}^{\lambda_0}]=
\mathrm{e}^{\lambda_0}\{
\tfrac{x}{4}(\partial_x[\phi_0])^2
+\tfrac{1}{x}(1-\mathrm{e}^{\lambda_0})
+\tfrac{x}{2}\mathrm{e}^{\lambda_0}V[\phi_0]\},\\
\label{eq:st nu}
\partial_x[\mathrm{e}^{\nu_0}]=
\mathrm{e}^{\nu_0}\{
\tfrac{x}{4}(\partial_x[\phi_0])^2
-\tfrac{1}{x}(1-\mathrm{e}^{\lambda_0})
-\tfrac{x}{2}\mathrm{e}^{\lambda_0}V[\phi_0]\},\\
\label{eq:st:wave}
\partial^2_x[\phi_0]
+(\tfrac{2}{x}+\tfrac{1}{2}\partial_x[\nu_0-\lambda_0])
\partial_x[\phi_0]
-\mathrm{e}^{\lambda_0}\delta_\phi V[\phi_0]=0,
\end{gather}
\end{subequations}
where we have written $\nu_0$, $\lambda_0$ and $\phi_0$ to represent the static background fields.

We will employ the simple shooting method~\cite[Section~7.3.1]{Stoer+Bulirsch:1993} on $\phi_0$ to numerically generate a solution to~\eqref{eq:static eqns}.
Initial values for $\exp(\nu_0)$, $\exp(\lambda_0)$, $\phi_0$ and $\partial_x[\phi_0]$ are given for small $x$ and~\eqref{eq:static eqns} are integrated outward using a fourth-order Runge-Kutta integration scheme~\cite[Section~7.2.1]{Stoer+Bulirsch:1993}.
Assuming that both of $\exp(\nu_0)$ and $\exp(\lambda_0)$ vanish as some positive power of $x$ as $x\rightarrow 0$, we find ($n_+:=2(\sqrt{1+k^2}+1+k^2)/k^2\in[2,\infty]$ and $x_k:=x/\sqrt{1+k^2}$)
\begin{equation}\label{eq:small}
\mathrm{e}^{\nu_0}\sim
F x_k^{n_+-2},\quad
\mathrm{e}^{\lambda_0}\sim
G n_+^2 x_k^{n_+},\quad
\phi_0\sim
-k (n_+-2)\ln(x_k)+B,
\end{equation}
independent of the values of $\tilde{m}$ and $\tilde{\kappa}$.
The parameterization in~\eqref{eq:small} is motivated by the analytic $\tilde{m}=\tilde{\kappa}=0$ solution, which from~\eqref{eq:Wyman small x} has $F=G=1$ and $B=0$.
(Note that when the potential vanishes the field equations are invariant under $\phi_0\rightarrow\phi_0+\text{constant}$.
This must be taken into account when applying the shooting method, however it is sufficient to check that $B=0$ and $F=G=1$ reproduces the analytically known solution.)

The small-$x$ behaviour of these solutions indicates that there is a curvature singularity at $x=0$, showing up in the Ricci scalar as $R\propto 1/x^4$.
Furthermore, by considering radial null geodesics emanating from $x=0$ (affinely parameterized by $\lambda$ and with tangent $u$), we find $\lambda^2R(u,u)\sim 1/n_+^2$; the finite limit of this as $\lambda\rightarrow 0$ indicates that there is a strong curvature singularity at $x=0$ by an argument of Clarke and Krolak (see for instance~\cite{Tipler+Clarke+Ellis:1980}).
For the solutions considered herein, it is straightforward to show that outgoing radial null geodesics are well-defined and reach $x=\infty$ in an infinite amount of affine time, as do outgoing radial timelike geodesics with sufficient energy.

The large-$x$ behaviour of $\phi_0$ is determined by requiring that the metric components be of the asymptotic Schwarzschild forms
\begin{equation}\label{eq:asmpt Sch}
\mathrm{e}^{\nu_0}\sim 1-\tfrac{1}{x},\quad
\mathrm{e}^{\lambda_0}\sim 1+\tfrac{1}{x}.
\end{equation}
Inserting these into~\eqref{eq:st:wave} we find (keeping only the asymptotically dominant terms and noting that we require that $\phi_0\rightarrow 0$ as $x\rightarrow\infty$)
\begin{equation}
\partial^2_x[\phi_0]
+\tfrac{2}{x}\partial_x[\phi_0]
-\tilde{m}^2\phi_0\sim 0.
\end{equation}
From this, the asymptotic form of $\phi_0$ is determined:
\begin{equation}
\phi_0\sim\begin{cases}
A/x& \text{for }\tilde{m}=0\\
A\mathrm{e}^{-\tilde{m}x}/x& \text{for }\tilde{m}\neq 0
\end{cases},
\end{equation}
and from~\eqref{eq:g to Z} and~\eqref{eq:Wyman large x} we know that for $\tilde{m}=\tilde{\kappa}=0$ we must find $A=k$.

Given input parameters $\tilde{m}$ and $\tilde{\kappa}$, the system~\eqref{eq:static eqns} has a unique solution once the initial data $k$, $F$, $G$ and $B$ are chosen.
Although the constant $F$ may be freely chosen by rescaling the time variable (and is chosen by requiring that $\exp(\nu_0)\rightarrow 1$ as $x\rightarrow\infty$), we must still determine $k$, $G$ and $B$.
For each value of $k$ that we consider, we avoid shooting on both of $G$ and $B$ by proceeding as follows: 
We choose values for $\tilde{m}$ and $\tilde{\kappa}$ and assign $F=G=1$ initially, shooting on $\phi_0$ to determine $B$.
This will result in a solution with asymptotic behaviour $\exp(\lambda_0)\sim (1+\alpha/x)$ and $\exp(\nu_0)\sim f_0(1-\alpha/x)$.
Re-scaling $x$ by $x\rightarrow \alpha x$ and renormalizing $\exp(\nu_0)$ results in a solution with the correct asymptotic form for the metric functions, while the $x\sim 0$ forms become 
\begin{equation}
\mathrm{e}^{\nu_0}\sim f_0^{-1} (\alpha x_k)^{n_+-2},\quad
\mathrm{e}^{\lambda_0}\sim n_+^2(\alpha x_k)^{n_+}.
\end{equation}
In addition, the constant $B$ determined by shooting is shifted by $B\rightarrow B-k(n_+-2)\ln(\alpha)$.
Most importantly, we have $\tilde{m}\rightarrow \alpha\tilde{m}$ and $\tilde{\kappa}\rightarrow \alpha^2\tilde{\kappa}$, and we cannot treat $\tilde{m}$ and $\tilde{\kappa}$ as fixed input parameters.
Note that any solution that satisfies $\phi_0\rightarrow 0$ as $x\rightarrow 0$ may be rescaled to one that has $F=G=1$, and therefore we are exploring the entire class of solutions in this manner.

\begin{figure}[h]
\begin{center}
\includegraphics[scale=1.0]{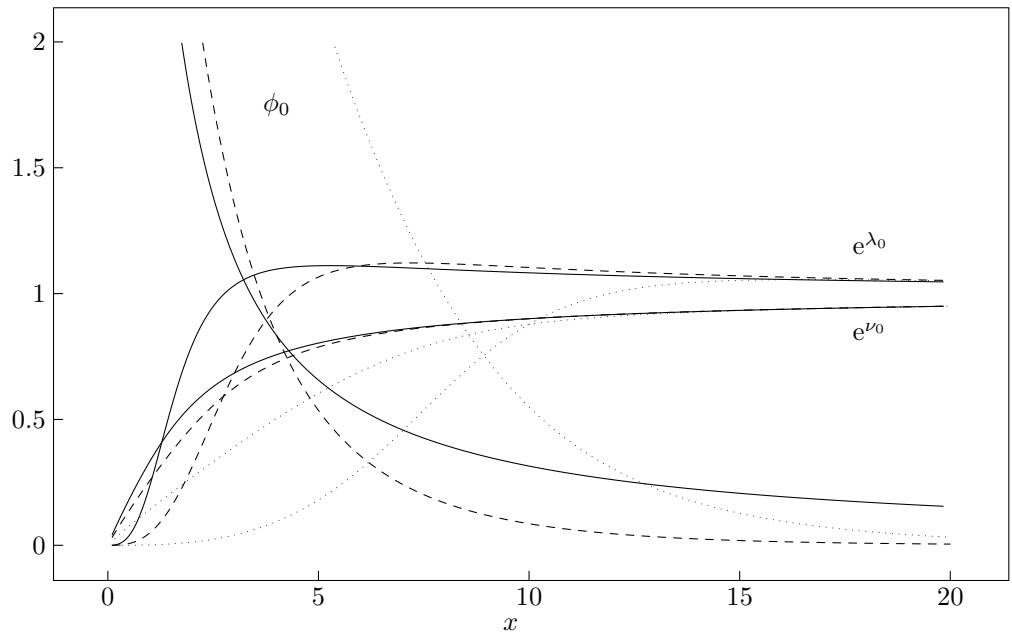}
\end{center}
\caption{
The background metric functions for three $k=3$ solutions: the solid line corresponds to the unstable $\tilde{m}=\tilde{\kappa}=0$ solution, the dashed line the unstable $\tilde{\kappa}=0$, $\tilde{m}\approx 0.210$ solution, and the dotted line the stable $\tilde{\kappa}=0$, $\tilde{m}\approx 0.210$ solution.
}
\label{fig:all}
\end{figure}

We have found that although $k$, $F$, $G$ and $B$ uniquely determine a solution of~\eqref{eq:static eqns} and $k$ and $B$ parameterize the space of solutions with asymptotic form given by~\eqref{eq:asmpt Sch}, for the $\tilde{\kappa}=0$ solutions this does not uniquely determine $\tilde{m}$.
We show in Figure~\ref{fig:all} two solutions with $k=3$ and $\tilde{m}\approx 0.210$ as well as the $k=3$, $\tilde{m}=\tilde{\kappa}=0$ solution.
From Figure~\ref{fig:kappa zero} we see the following feature: above a minimum $\tilde{m}$ there is a stable and an unstable solution labeled by the same values of $k$ and $\tilde{m}$.
(A similar result holds for the $\tilde{m}=0$ solutions with $\tilde{\kappa}\neq 0$ not considered in this work.)

\section{Dynamical Stability}
\label{sect:instability}

We turn to the investigation of the dynamical stability of these spacetimes, \textit{i.e.}, whether they are stable against linear perturbations.
Two different methods will be employed: a variational approach which casts the equation for the perturbative modes into the form of a stationary Schr\"{o}dinger equation and uses a one-parameter family of wave functions to put a least upper bound on the lowest eigenvalue, and a shooting method to numerically generate both the lowest energy mode as well as the eigenvalue.

In either case we make use of the single Fourier-mode expansions of the perturbing fields as $\nu(t,x)=\nu_0(x)+\nu_1(x)\cos(\tilde{\omega} t)$, $\lambda(t,x)=\lambda_0(x)+\lambda_1(x)\cos(\tilde{\omega} t)$ and $\phi(t,x)=\phi_0(x)+\phi_1(x)\cos(\tilde{\omega} t)/x$.
The first-order perturbation equations from~\eqref{eq:feq used} for the metric functions give
\begin{subequations}
\begin{equation}
\lambda_1=\tfrac{1}{2}\partial_x[\phi_0]\phi_1,\quad
\partial_x[\nu_1]=-\tfrac{1}{2}\partial^2_x[\phi_0]\phi_1
-\tfrac{1}{x}\partial_x[\phi_0]\phi_1
+\tfrac{1}{2}\partial_x[\phi_0]\partial_x[\phi_1],
\end{equation}
and the equation for $\phi_1$
\begin{equation}\label{eq:phi 1}
\partial_x^2[\phi_1]
+\tfrac{1}{2}\partial_x[\nu_0-\lambda_0]\partial_x[\phi_1]
-\mathrm{e}^{\lambda_0-\nu_0}V_0[x]\phi_1
+\tilde{\omega}^2\mathrm{e}^{\lambda_0-\nu_0}\phi_1=0,
\end{equation}
where
\begin{equation}
\begin{split}
V_0[x]:=\mathrm{e}^{\nu_0-\lambda_0}\bigl\{
&\tfrac{1}{2x}\partial_x[\nu_0-\lambda_0]
-\tfrac{1}{2}(\partial_x[\phi_0])^2
-\tfrac{1}{4}x(\partial_x[\phi_0])^2\partial_x[\nu_0-\lambda_0]\\
&+\mathrm{e}^{\lambda_0}(\tilde{m}^2+3\tilde{\kappa}\phi_0^2)
+x\mathrm{e}^{\lambda_0}\partial_x[\phi_0]
(\tilde{m}^2\phi_0+\tilde{\kappa}\phi_0^3)\bigr\}.
\end{split}
\end{equation}
\end{subequations}

\begin{figure}[h]
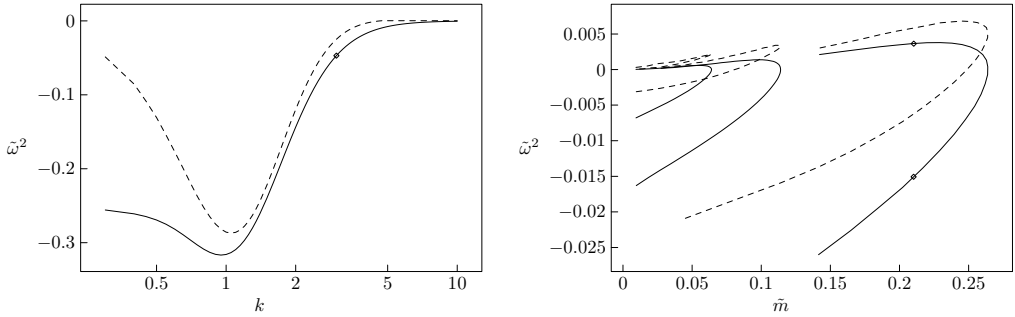

\begin{center}
\begin{subfigure}[$\tilde{m}=\tilde{\kappa}=0$.]
{\includegraphics[scale=0.7]{Omega-k-mu-0-kappa-0.ps}
\label{fig:mkappa zero}}
\end{subfigure}
\begin{subfigure}[$\tilde{\kappa}=0$ for (from right to left) $k=3$, $k=4$, $k=5$.]
{\includegraphics[scale=0.7]{Omega-mu-kappa-0.ps}
\label{fig:kappa zero}}
\end{subfigure}
\end{center}
\caption{In both figures the solid line indicates the exactly determined eigenvalue from numerically solving~\eqref{eq:phi 1} and the dashed line the variational bound generated from~\eqref{eq:var bound}.  
The diamonds indicate the solutions studied in Figures~\ref{fig:all} and~\ref{fig:Vx}.
}
\label{fig:Omegas}
\end{figure}

From the small-$x$ behaviour of the background fields~\eqref{eq:small} we find
\begin{equation}
\partial_x^2[\phi_1]-(1/x)\partial_x[\phi_1]+(1/x^2)\phi_1\sim 0,
\end{equation}
from which we have that $\phi_1\sim x(a+b\ln(x))$.
We will restrict ourselves to $b=0$ so that the perturbation is finite at $x=0$, however all of the solutions considered here appear unstable against singular perturbations with $b\neq 0$ and therefore $\ln(x)$ behaviour at the centre of symmetry.
Once the arbitrary amplitude $a$ is chosen,~\eqref{eq:phi 1} is integrated numerically, varying the value of $\tilde{\omega}^2$ and requiring that $\phi_1\rightarrow 0$ as $x\rightarrow \infty$.
In Figure~\ref{fig:mkappa zero}, we display the results of this procedure for the $\tilde{m}=\tilde{\kappa}=0$ solutions, giving the value of $\tilde{\omega}^2$ for lowest energy mode.
As we see, $\tilde{\omega}^2<0$, indicating that these solutions are generically unstable.
As discussed earlier, there are two solutions with given values of $k$ and $\tilde{m}$ and, as shown in Figure~\ref{fig:kappa zero}, there is an unstable branch ($\tilde{\omega}^2<0$) and stable branch ($\tilde{\omega}^2>0$).

In generating the perturbative ground state and eigenvalue for the $\tilde{m}\neq 0$ solutions, the presence of an exponentially diverging solution of~\eqref{eq:feq used} does not allow us to generate the background solution to arbitrarily large $x$.
However we shall see from~\eqref{eq:asympt V} and~\eqref{eq:Sch 2} that for large enough $\rho$ (and $\tilde{\omega}^2<\tilde{m}^2$) we will always have $\partial_x^2[\phi_1]>0$, and therefore if $\partial_x[\phi_1]>0$ then the perturbation \textit{must} diverge.
When shooting on $\tilde{\omega}^2$ to determine the ground state, it is therefore sufficient to check whether $\phi_1$ passes through zero (in which case $\tilde{\omega}^2$ is too large) or $\phi_1\partial_x[\phi_1]>0$ for large enough $x$ that $V-\tilde{\omega}^2>0$ (in which case $\tilde{\omega}^2$ is too small).

\section{The Variational method}
\label{sect:var}

We now consider the variational method of generating an upper bound on $\tilde{\omega}^2$, introduced by Jetzer and Scialom~\cite{Jetzer+Scialom:1992}.
Although we find no fault with the method in principle, we show that they made a technical error in its application.
Using their erroneous bound, one would find that these solutions are generically unstable, whereas the direct numerical determination of the ground state and eigenvalue in Section~\ref{sect:instability} contradicts their bound and furthermore indicates that there are solutions to the massive scalar field equations that are perturbatively stable (as shown in Figure~\ref{fig:kappa zero}).

Following~\cite{Jetzer+Scialom:1992} we introduce a radial coordinate $\rho$ that eliminates the second term in~\eqref{eq:phi 1}
\begin{subequations}\label{eq:rho equation}
\begin{equation}
\rho=\int_0^x dx\;\mathrm{e}^{-\tfrac{1}{2}(\nu_0-\lambda_0)},
\end{equation}
chosen so that $\rho=0$ at $x=0$.
 From~\eqref{eq:small} and~\eqref{eq:asmpt Sch} respectively we find that
\begin{equation}
\rho\sim 
\begin{cases}
\sqrt{G/F} x^2 n_+/(2\sqrt{1+k^2})&\text{as }x\rightarrow 0\\
x+\ln(x)+\mathrm{constant} &\text{as }x\rightarrow \infty\\
\end{cases};
\end{equation}
\end{subequations}
$\rho$ is numerically determined by making use of the $x\rightarrow 0$ result and integrating outward.

This coordinate transforms~\eqref{eq:phi 1} to the stationary Schr\"{o}dinger-like equation
\begin{equation}\label{eq:Sch 1}
-\frac{\partial^2\phi_1}{\partial\rho^2}+V[x(\rho)]\phi_1=\tilde{\omega}^2\phi_1.
\end{equation}
We consider perturbations in the real Hilbert space $L^2(d\rho,(0,\infty))$,
\textit{i.e.}, real functions on the half-line with the norm
\begin{equation}
(\phi,\phi^\prime)=\int_0^\infty d\rho\,
\phi(\rho) \phi^\prime(\rho)
=\int_0^\infty dx\,
\mathrm{e}^{-\tfrac{1}{2}(\nu_0-\lambda_0)}
\phi[\rho(x)] \phi^\prime[\rho(x)].
\end{equation}
From the asymptotic and small-$x$ forms given earlier, we find
\begin{equation}\label{eq:asympt V}
V[\rho]\xrightarrow{\rho\rightarrow 0}-1/(4\rho^2),\quad
V[\rho]\xrightarrow{\rho\rightarrow\infty}
\begin{cases}
-1/\rho^3&\text{for }\tilde{m}=\tilde{\kappa}=0\\
3\tilde{\kappa}A^2/\rho^2&
\text{for }\tilde{m}=0,\quad \tilde{\kappa}\neq 0\\
\tilde{m}^2&\text{for }\tilde{m}\neq 0
\end{cases},
\end{equation}
and we therefore introduce
\begin{equation}
H_0=-\partial_\rho^2-\frac{1}{4\rho^2},\quad
\tilde{V}[x(\rho)]=\frac{1}{4\rho^2}+V[x(\rho)],
\end{equation}
so that~\eqref{eq:Sch 1} is written as
\begin{equation}\label{eq:Sch 2}
H_0\phi_1+\tilde{V}[x(\rho)]\phi_1=\tilde{\omega}^2\phi_1.
\end{equation}

From this form it is worthwhile to make a few observations.
If we consider regions where $\tilde{V}\approx \tilde{V}_0=\text{constant}$, then for $\tilde{\omega}^2>\tilde{V}_0$ we have the oscillating Bessel function solutions (defining $\tilde{\Omega}:=\sqrt{\lvert\tilde{\omega}^2-\tilde{V}_0\rvert}$)
\begin{equation}\label{eq:constant}
\phi_1\approx
A\sqrt{\rho}J_0(\tilde{\Omega}\rho)
+B\sqrt{\rho}Y_0(\tilde{\Omega}\rho),
\end{equation}
and for $\tilde{\omega}^2<\tilde{V}_0$ replace $J_0$ and $Y_0$ with $I_0$ and $K_0$ to find the exponentially increasing and damped solutions respectively.
From this we deduce that $\tilde{\omega}^2\le \lim_{x\rightarrow\infty}\tilde{V}$ since otherwise from~\eqref{eq:constant} $\phi_1$ would not vanish as $x\rightarrow\infty$, and from~\eqref{eq:asympt V} we therefore find that the solutions with $\tilde{m}=0$ can have no stable modes; we will not consider stationary perturbations ($\tilde{\omega}=0$) in this work.
This has been verified for the $\tilde{m}=\tilde{\kappa}=0$ solutions as shown in Figure~\ref{fig:mkappa zero} as well as for the $\tilde{m}=0$ and $\tilde{\kappa}\neq 0$ solutions.
(Note that one must be careful in the latter case since for arbitrarily small and positive $\tilde{\omega}^2$, $\phi_1$ will begin oscillating at arbitrarily large $x$ even though it will appear to have exponential behaviour prior to that.)
This does allow the possibility that the $\tilde{m}\neq 0$ solutions are stable, and from Figure~\ref{fig:kappa zero} we have seen that there is a branch of solutions for which this is realized.

From the work of Narnhofer~\cite{Narnhofer:1974} we know that the operator $H_0$ is not self-adjoint on the intersecting domain $\mathcal{D}(-\partial_\rho^2)\cap\mathcal{D}(-1/(4\rho^2))$, with deficiency indices $(1,1)$.
A family of self-adjoint extensions is determined by extending this domain to include the solutions of 
\begin{subequations}\label{eq:psi prop}
\begin{equation}
-\partial_\rho^2[\psi_{\pm i}]-1/(4\rho^2)\psi_{\pm i}=\pm i\psi_{\pm i}, 
\end{equation}
which in this case are 
\begin{equation}
\psi_i=\sqrt{\rho}H^{(1)}_0(\rho\mathrm{e}^{i\pi/4}),\quad
\psi_{-i}=\bar{\psi}_i=\sqrt{\rho}H^{(2)}_0(\rho\mathrm{e}^{-i\pi/4}),
\end{equation}
\end{subequations}
where $H_0^{(1)}$ and $H_0^{(2)}$ are the zeroth-order Hankel functions of the first and second kind respectively.

The self-adjoint extensions are parameterized by the real angle $\tau$, and extend the operator to act on linear combinations of the real functions
\begin{equation}\label{eq:Psitau}
\Psi_\tau:=\tfrac{1}{2}
(\mathrm{e}^{i\tau}\psi_i+\mathrm{e}^{-i\tau}\psi_{-i})
=\cos(\tau)\Re(\psi_i)-\sin(\tau)\Im(\psi_i),
\end{equation}
where $\Re$ and $\Im$ represent the real and imaginary parts respectively.
(Note that we consider self-adjoint extensions on a real domain only since the scalar field is real.)
We write the operator $\bar{H}_{0,\tau}$ which acts like $H_0$ on the extended domain $\mathcal{D}(\bar{H}_{0,\tau})
=\mathcal{D}(-\partial_\rho^2)\cap
\mathcal{D}(-1/(4\rho^2))
+\{\Psi_\tau \}$, and using~\eqref{eq:psi prop} acts on $\Psi_\tau$ as
\begin{equation}\label{eq:HPsitau}
\bar{H}_{0,\tau}\Psi_\tau=\tfrac{1}{2}i
(\mathrm{e}^{i\tau}\psi_i-\mathrm{e}^{-i\tau}\psi_{-i})
=-\sin(\tau)\Re(\psi_i)-\cos(\tau)\Im(\psi_i).
\end{equation}
We will require the following integrals:
\begin{subequations}\label{eq:params}
\begin{gather}
\int_0^\infty d\rho\;\Re(\psi_i)^2
=\int_0^\infty d\rho\;\Im(\psi_i)^2
=:a\approx 0.1592,\\
\int_0^\infty d\rho\;\Re(\psi_i)\Im(\psi_i)
=:b\approx -0.1013,
\end{gather}
\end{subequations}
and using~\eqref{eq:Psitau} and~\eqref{eq:HPsitau} we find
\begin{equation}\label{eq:inn results}
(\Psi_\tau,\Psi_\tau)
=(a-b\sin(2\tau)),\quad
(\Psi_\tau,\bar{H}_{0,\tau}\Psi_\tau)
=-b\cos(2\tau).
\end{equation}
This latter result corrects the error in the work of Jetzer and Scialom who have assumed that it vanishes (see the paragraph following Equation~$(28)$ in~\cite{Jetzer+Scialom:1992}).

From the asymptotic forms of the Hankel functions, we find the behaviour of $\Psi_\tau$ for large and small $\rho$ respectively to be
\begin{subequations}
\begin{align}
\label{eq:largez}
\Psi_\tau&\sim
\sqrt{2/\pi}\mathrm{e}^{-\rho/\sqrt{2}}
\sin(\rho/\sqrt{2}+\pi/8+\tau),\\
\label{eq:Psi rho 0}
\Psi_\tau&\sim 
\sqrt{\rho}\cos(\tau)/2
-(2\sqrt{\rho}/\pi)\sin(\tau)(\gamma_{\textsc{e}}+\ln(\rho/2)).
\end{align}
\end{subequations}
At this point we note that only the $\tau=0$ extension (chosen implicitly in~\cite{Jetzer+Scialom:1992}) corresponds to a perturbative field that is finite at $\rho=0$ and, consistent with the choice made in Section~\ref{sect:instability}, we restrict ourselves to this case in what remains.

From the form of the differential operator $\bar{H}_{0,0}$, for any two functions $\psi,\psi^\prime \in\mathcal{D}(\bar{H}_{0,0})$ we define $\psi_\beta(\rho):=\psi(\beta\rho)$, and it is straightforward to prove the scaling property: $(\psi^\prime_\beta,\bar{H}_{0,0}\psi_\beta)=\beta(\psi^\prime,\bar{H}_{0,0}\psi)$.
We therefore consider the collection of functions
\begin{equation}
\Psi_{0,\beta}(\rho):=\Psi_0(\beta\rho),
\end{equation}
as the variational family, varying $\beta$ to get a least upper bound on the ground state energy.
Using this scaling property and the results~\eqref{eq:params} and~\eqref{eq:inn results}, we find
\begin{equation}
(\Psi_{0,\beta},\bar{H}_{0,0}\Psi_{0,\beta})
=-b\beta,\quad
(\Psi_{0,\beta},\Psi_{0,\beta})
=a/\beta,
\end{equation}
and taking the expectation value of~\eqref{eq:Sch 2} leads to the bound
\begin{equation}\label{eq:var bound}
\tilde{\omega}^2\leq
\beta((\Psi_{0,\beta},\tilde{V}\Psi_{0,\beta})-b\beta)/a
=:\mathrm{max}(\tilde{\omega}^2).
\end{equation}

\begin{figure}[h]
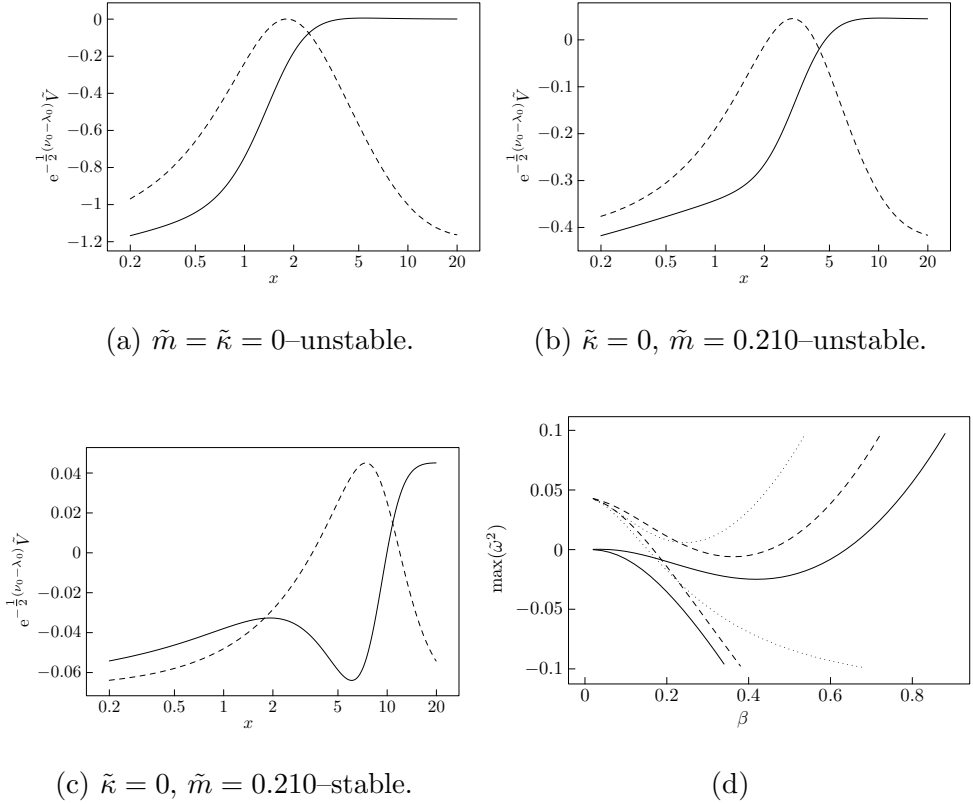

\begin{center}
\begin{subfigure}[$\tilde{m}=\tilde{\kappa}=0$--unstable.]
{\includegraphics[scale=0.65]{niftyPlot.0.ps}}
\end{subfigure}
\begin{subfigure}[$\tilde{\kappa}=0$, $\tilde{m}=0.210$--unstable.]
{\includegraphics[scale=0.65]{niftyPlot.1.ps}}
\end{subfigure}
\begin{subfigure}[$\tilde{\kappa}=0$, $\tilde{m}=0.210$--stable.]
{\includegraphics[scale=0.65]{niftyPlot.2.ps}}
\end{subfigure}
\begin{subfigure}[]
{\includegraphics[scale=0.7]{ScanBeta.ps}
\label{fig:scan beta}}
\end{subfigure}
\end{center}
\caption{On the first three figures appear the effective potential (solid line) and the perturbative ground state (dotted line) for the solutions that appear in Figure~\ref{fig:all}.
The final figure shows the behaviour of $\mathrm{max}(\tilde{\omega}^2)$~\eqref{eq:var bound} with respect to $\beta$ (labeled as in Figure~\ref{fig:all}): the curves with an obvious minimum correspond to the correct form of the bound~\eqref{eq:var bound}, and those that diverge negatively are derived from~\eqref{eq:var bound} by dropping the second term and corresponds to that of Jetzer and Scialom.
Note that this final form clearly contradicts the results shown in Figure~\ref{fig:Omegas}.
}
\label{fig:Vx}
\end{figure}

The inequality~\eqref{eq:var bound} should be compared with Equation~$(30)$ of~\cite{Jetzer+Scialom:1992} which is lacking the $-b\beta^2/a$ contribution.
As we see from Figure~\ref{fig:scan beta}, this term is necessary for the consistency of the variational result with the numerically generated eigenvalue, as it dominates for large $\beta$.
Na\"{\i}vely we expect that very small values of $\beta$ will result in a bound of $\tilde{\omega}^2\le \tilde{m}^2$ since in these cases the bulk of the support of $\Psi_{0,\beta}$ is shifted to large $\rho$ and the asymptotic behaviour of the $\tilde{V}$~\eqref{eq:asympt V} in the expectation value dominates the integral.
This is indeed what we see from Figure~\ref{fig:scan beta}, and is consistent with the discussion following~\eqref{eq:Sch 2}.
In Figure~\ref{fig:Vx} we also display the effective potential and perturbative ground state for the three spacetimes in Figure~\ref{fig:all}.
We see from Figure~\ref{fig:kappa zero} that the bound is not very useful for considering background spacetimes with $\tilde{m}\neq 0$ and `large' values of $k$, as it gives a positive upper bound for unstable solutions.

\section{Discussion}
\label{sect:discussion}

The existence of perturbatively stable, nakedly singular spacetimes is interesting, and perhaps surprising if one is a proponent of the cosmic censorship hypothesis.
We should make some cautionary remarks though, since we have in no way demonstrated a violation of this hypothesis.
To begin with, one really should examine non-spherical perturbations of these spacetimes along the lines of the stability proof of the Schwarzschild solution~\cite{Regge+Wheeler:1957,Vishveshwara:1970} and developed further in~\cite{Kojima+Yoshida+Futamase:1991}.
The existence of non-spherically symmetric unstable modes would indicate that the existence of any asphericity in a  collapsing system would drive it away from the nakedly singular configuration, and would not be accessible from the study of the spherically symmetric collapse problem.

However if there continues to be an absence of unstable modes, then we are in a somewhat uncomfortable situation.
We know from the work of Christodoulou~\cite{Christodoulou:1994} that naked singularity formation in the collapse of massless scalar fields is possible, however only with highly tuned initial data.
It is likely then that massive fields can also form naked singularities, since the behaviour of the fields at the singular point is independent of the scalar field potential.
If the stable branch can be reached from the collapse of suitably regular initial data, then we would na\"{\i}vely expect that there are choices of initial data that end up `near' the nakedly singular spacetime, and thus the initial data is less special in this case.
At the very least this is motivation to investigate this system further. 

In addition to the solutions considered herein, there are also those with vanishing and negative asymptotic Schwarzschild mass with the same behaviour at the centre of symmetry, as well as those with non-vanishing $\tilde{\kappa}$.
These will be considered in more detail in the future.

\ack

The authors thank the Natural Sciences and Engineering Research Council of Canada, the Walter~C.~Sumner Foundation, the Government of Ontario (OGS), and the Department of Physics at the University of Toronto for financial support.

\appendix

\section{{W}yman's Solution}
\label{sect:Wyman}

Here we include some analytic results on exact solutions with $V[\phi]=0$ attributed to Wyman~\cite{Wyman:1981}.
The equation for the scalar field~\eqref{eq:st:wave} is integrated to find (choosing here $k>0$ so that $\phi_0$ is asymptotically positive and decreasing to zero)
\begin{equation}
\partial_x[\phi_0]=-k\mathrm{e}^{-\tfrac{1}{2}(\nu_0-\lambda_0)}/x^2,
\end{equation}
and integrating $R_{00}=\exp(\nu_0-\lambda_0)(\partial_x^2[\nu_0]/2+(\partial_x[\nu_0])^2/4-\partial_x[\nu_0]\partial_x[\lambda_0]/4+\partial_x[\nu_0]/x)=0$, results in (introducing the integration constant $\alpha$)
\begin{equation}\label{eq:integ R00}
\partial_x[\nu_0]=\alpha k\mathrm{e}^{-\tfrac{1}{2}(\nu_0-\lambda_0)}/x^2
=-\alpha\partial_x[\phi_0],\quad\text{which leads to}\quad
\nu_0=-\alpha\phi_0,
\end{equation}
where a possible constant of integration has been set equal to zero by requiring that $\{\nu_0,\phi_0\}\rightarrow 0$ as $x\rightarrow \infty$.
Assuming the asymptotic forms of $\nu_0$ and $\lambda_0$ corresponding to positive and negative asymptotic Schwarzschild masses $\nu_0\sim\mp 1/x$ and $\lambda_0\sim\pm 1/x$, we find that $\alpha=\pm 1/k$.

Introducing $Z:=\pm k\exp(-\nu_0/2+\lambda_0/2)/(2x)$ we derive the ``transition equation''
\begin{equation}\label{eq:transition}
x\partial_x[Z]=Z(Z^2-2Z/k-1)
=Z(Z-Z_+)(Z-Z_-),
\end{equation}
where $Z_{\pm}:=(1\pm\sqrt{1+k^2})/k$ and we will make use of $n_\pm:=Z_\pm(Z_\pm-Z_\mp)=\pm 2(\sqrt{1+k^2}\pm 1\pm k^2)/k^2$.
From its definition we require that $Z\rightarrow 0$ as $x\rightarrow 0$ from below (above), and from~\eqref{eq:transition} we see that $Z\rightarrow Z_\pm$ as $x\rightarrow 0$ for the positive and negative asymptotic Schwarzschild mass solutions respectively.
The implicit solution to~\eqref{eq:transition} with integration constant chosen so that $Z\rightarrow\pm k/(2x)$ as $x\rightarrow\infty$ is
\begin{equation}\label{eq:implicit}
x=\pm\frac{k}{2Z}
( 1-Z/Z_+)^{1/n_+}
( 1-Z/Z_-)^{1/n_-}.
\end{equation}
In terms of $Z$ we have
\begin{subequations}\label{eq:g to Z}
\begin{gather}
\mathrm{e}^{\nu_0}=\frac{k^2}{4x^2Z^2}(1+2Z/k-Z^2),\quad
\mathrm{e}^{\lambda_0}=(1+2Z/k-Z^2),\\
\phi=\mp k\ln
\Bigl[\frac{k^2}{4x^2Z^2}(1+2Z/k-Z^2)\Bigr],
\end{gather}
\end{subequations}
and the large-$x$ behaviour of~\eqref{eq:implicit} we have
\begin{equation}\label{eq:Wyman large x}
Z\sim\pm\frac{k}{2x}(1\pm 1/x +o(1/x)),
\end{equation}
which is consistent with the required behaviour $\exp(\nu_0)\sim 1\mp 1/x$ and $\exp(\lambda_0)\sim 1\pm 1/x$.
The small-$x$ expansion gives (defining $x_k:=x/\sqrt{1+k^2}$)
\begin{equation}
Z\sim Z_\pm( 1-n_\pm x_k^{n_\pm}),
\end{equation}
which leads via~\eqref{eq:g to Z} to
\begin{subequations}
\begin{gather}\label{eq:Wyman small x}
\mathrm{e}^{\nu_0}\sim
x_k^{n_\pm-2},\quad
\mathrm{e}^{\lambda_0}\sim 
n_\pm^2x_k^{n_\pm},\quad
\phi_0\sim\mp k (n_{\pm}-2)\ln(x_k).
\end{gather}
\end{subequations}



\providecommand{\bysame}{\leavevmode\hbox to3em{\hrulefill}\thinspace}

\end{document}